\documentclass[journal=apchd5,manuscript=article]{achemso}

\usepackage[utf8]{inputenc}
\usepackage[T1]{fontenc}
\usepackage{mathptmx}
\usepackage{soul}
\usepackage{etoolbox}
\usepackage{xcolor}
\usepackage{graphicx}
\usepackage{dcolumn}
\usepackage{bm}

\usepackage{soul}  
\usepackage{lineno}

\usepackage{siunitx} 
\usepackage[version=3]{mhchem} 

\author{Tecla Gabbrielli}
 \affiliation{CNR-INO -- Istituto Nazionale di Ottica, Via Carrara, 1 -- 50019 Sesto Fiorentino FI, Italy}
\alsoaffiliation{LENS -- European Laboratory for Non-Linear Spectroscopy, Via Carrara, 1 -- 50019 Sesto Fiorentino FI, Italy}
\altaffiliation{These authors equally contributed to this work.}
\email{tecla.gabbrielli@ino.cnr.it}
\author{Chenghong Zhang}%
\altaffiliation{These authors equally contributed to this work.}
 \affiliation{CNR-INO -- Istituto Nazionale di Ottica, Via Carrara, 1 -- 50019 Sesto Fiorentino FI, Italy}
\alsoaffiliation{LENS -- European Laboratory for Non-Linear Spectroscopy, Via Carrara, 1 -- 50019 Sesto Fiorentino FI, Italy}
\author{Francesco Cappelli}
\affiliation{CNR-INO -- Istituto Nazionale di Ottica, Via Carrara, 1 -- 50019 Sesto Fiorentino FI, Italy}
\alsoaffiliation{LENS -- European Laboratory for Non-Linear Spectroscopy, Via Carrara, 1 -- 50019 Sesto Fiorentino FI, Italy}
\author{Iacopo Galli}
\affiliation{CNR-INO -- Istituto Nazionale di Ottica, Via Carrara, 1 -- 50019 Sesto Fiorentino FI, Italy}
\alsoaffiliation{LENS -- European Laboratory for Non-Linear Spectroscopy, Via Carrara, 1 -- 50019 Sesto Fiorentino FI, Italy}
\author{Andrea Ottomaniello}
\affiliation{Center for Materials Interfaces, Istituto Italiano di Tecnologia, Via R. Piaggio, 34 -- 56025 Pontedera, PI, Italy}
\author{Jérôme Faist}
\affiliation{Institute for Quantum Electronics, ETH Z\"{u}rich, Z\"{u}rich, Switzerland}
\author{Alessandro Tredicucci}
\affiliation{Dipartimento di Fisica, Università di Pisa, Largo B. Pontecorvo 3, 56127 Pisa, Italy}
\alsoaffiliation{Laboratorio NEST, CNR -- Istituto Nanoscienze, Piazza San Silvestro 12 --56127  Pisa, Italy}
\author{Alessandro Pitanti}
\affiliation{Dipartimento di Fisica, Università di Pisa, Largo B. Pontecorvo 3, 56127 Pisa, Italy}
\alsoaffiliation{Laboratorio NEST, CNR -- Istituto Nanoscienze, Piazza San Silvestro 12 --56127  Pisa, Italy}
\author{Paolo De Natale}
\affiliation{CNR-INO -- Istituto Nazionale di Ottica, Via Carrara, 1 -- 50019 Sesto Fiorentino FI, Italy}
\alsoaffiliation{LENS -- European Laboratory for Non-Linear Spectroscopy, Via Carrara, 1 -- 50019 Sesto Fiorentino FI, Italy}
\author{Simone Borri}
\affiliation{CNR-INO -- Istituto Nazionale di Ottica, Via Carrara, 1 -- 50019 Sesto Fiorentino FI, Italy}
\alsoaffiliation{LENS -- European Laboratory for Non-Linear Spectroscopy, Via Carrara, 1 -- 50019 Sesto Fiorentino FI, Italy}
\author{Paolo Vezio}
\affiliation{LENS -- European Laboratory for Non-Linear Spectroscopy, Via Carrara, 1 -- 50019 Sesto Fiorentino FI, Italy}
\alsoaffiliation{Dipartimento di Fisica e Astronomia e Università di Firenze, Via Sansone 1 -- 50019 Sesto
Fiorentino FI, Italy }
\title
  {Bridging mid and near infrared by combining optomechanics and self mixing}


\begin{document}

\begin{abstract}
This work describes a self-mixing-assisted optomechanical platform for transferring information between near- and mid-infrared radiation. In particular, the self-mixing signal of a mid-infrared quantum cascade laser is used to detect the oscillation of a membrane driven by light-induced forces exerted by a near-infrared excitation beam, which is amplitude-modulated at the membrane resonance frequency. This technique benefits from spectral broadness and, therefore, can link different spectral regions from both the excitation and probe sides. This versatility can pave the way for future applications of this self-mixing-assisted optomechanical platform in communication and advanced sensing systems.
\end{abstract}

\section{Introduction}

Self-mixing (SM) detection is a homodyne optical feedback interferometry technique where the laser is used both as source and detector~\cite{Lang:1980,Giuliani:2002,taimre2015}. 
In SM-based schemes, the light emitted by the laser source is back-reflected via the target (e.g., a membrane or a mirror) to one of the laser facets. The back-reflected light, re-injected into the laser active region, interferes with the intracavity optical field with a phase depending on the target position~\cite{Lang:1980}. This injection alters the laser working point parameters, such as intracavity optical power and laser voltage drop. This technique enables the extraction of target information by directly monitoring the voltage signal measured at the laser terminals~\cite{Bertling:2013} or the laser output power from the other facet. \\ 
Quantum cascade lasers (QCL)~\cite{Faist:1994} are particularly well-suited to be used in SM setups. Similarly to bipolar semiconductor lasers, they are highly sensitive to optical feedback. On the other hand, since the laser transition takes place between the sublevels of the conduction band created by the semiconductor layers' heterostructure, the laser transition lifetime is very short ($< 1$~ps)~\cite{Faist:2013bo}. For this reason, QCLs can be modulated at high speed (GHz and above)~\cite{Hinkov:2016,Hinkov:2019} and, in principle, are also able to detect SM signals within the same bandwidth. As a consequence, QCLs have been successfully employed for realizing highly integrated sensitive sensors, as feedback-induced variations in voltage or output power can carry precise displacement and/or optical path information~\cite{Ottomaniello:19,Vezio:2025}. 
Over the years, these SM-based compact sensors have been extensively applied to a wide range of applications, including characterization of laser linewidth and $\alpha$ factor~\cite{vonStaden:06,Kumazaki_2008,cardilli2016linewidth}, displacement sensors~\cite{leng2011demonstration}, gas sensing and imaging~\cite{dean2011terahertz,dean2013coherent}. 
\\ Recent studies have explored similar configurations involving QCL self-mixing with suspended membranes, demonstrating the feasibility of hybrid photonic-mechanical platforms~\cite{Vezio:2025,Ottomaniello:19}.
{However, these previous studies primarily focused on stationary photothermal effects in systems where membrane motion was piezoelectrically driven and only perturbed by external radiation. In contrast, here we explore a different regime in which dynamic light-induced forces—originating from both radiation pressure and photothermal effects~\cite{Metzger:2008}—serve as the primary driving mechanism for membrane oscillation. Specifically, we demonstrate a fully optical actuation scheme in which a trampoline membrane is driven by amplitude modulation of a near-infrared (near-IR) excitation laser and monitored via self-mixing interferometry using a mid-infrared (mid-IR) QCL.

In other words, we test and demonstrate the ability to encode the membrane oscillation signal, induced by the near-IR excitation beam, in the mid-IR QCL via the SM coupling. 
This means that the membrane, combined with SM, actively enables the transfer of information, i.e., a communication between the two beams at different infrared wavelengths. 
Moreover, by demonstrating that the light-induced force exerted by the excitation beam is the phenomenon responsible for the membrane oscillation, we prove that this SM-assisted optomechanical platform benefits from being, in principle,  broadband in terms of excitation wavelength.
\\ On one hand, this platform can be used as a communication gate between the mid-IR and other spectral ranges. On the other hand, the universal nature of the SM effect guarantees wavelength independence for the probe source. This further expands the applicability of the proposed platform across most of the frequency spectral range.  \\  
\section{Experimental Setup and Methodology Description}
\label{sec:Exp_setup}
\begin{figure*}[htb!]
    \centering
    \includegraphics[width=0.8\linewidth]{ 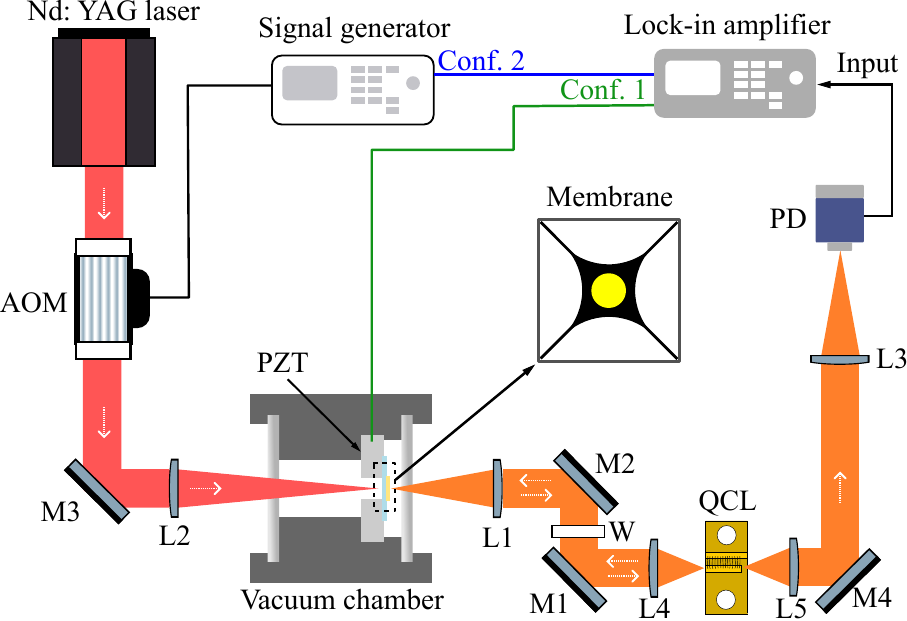}
    \caption{ Schematic of the SM setup. M: mirror, W: window, PD: photodetector, L: lens, PZT: piezoelectric actuator, AOM: acousto-optic modulator. The configuration used to characterize the membrane resonance frequency by acting on the piezo, namely \textit{Conf. 1} is depicted in green. The configuration where the membrane is excited via the AM-modulated near-IR radiation is depicted in blue, namely, \textit{Conf. 2}.\label{fig:exp_setup}}
\end{figure*}
The experimental setup is shown in Fig.~\ref{fig:exp_setup}. 
The SM signal generated in a Fabry-Pérot QCL emitting at \SI{4.5}{\micro m} is used to probe and monitor the optomechanical oscillation induced in a trampoline membrane via a near-IR excitation beam.
\\ In particular, the trampoline membrane is a silicon nitride (Si$_3$N$_4$) membrane with a metal coating (Cr/Au 5-30 nm coating).  More details about the membrane design and fabrication are available in ref. \cite{Vezio:2025}. The membrane is fixed on a pass-through-holed piezoelectric actuator (PZT). The PZT can actuate the membrane oscillation, e.g., for calibration purposes. The membrane is housed in a room-temperature vacuum chamber featuring two selected windows that maintain transparency while enabling simultaneous transmission of the impinging laser beams. In detail, a 1-mm-thick cyclic-olefin copolymer optical window is used on the near-IR excitation beam side and a CaF$_{2}$ window on the mid-IR radiation side. The vacuum chamber is pumped at $ \sim 10^{-3}$~mbar. For alignment purposes, the membrane platform (membrane, vacuum chamber, PZT; see Fig.~\ref{fig:exp_setup}) is mounted on a 3-axis stage.\\ On one side, the membrane is illuminated by the near-IR excitation beam (left side in Fig.~\ref{fig:exp_setup}). This beam is obtained by deflecting the radiation of an Nd:YAG laser, emitting at \SI{1064}{nm}, via an acousto-optic modulator (AOM). The AOM deflects the laser beam and shifts the laser frequency by 245 MHz through a radio frequency (RF) injection. The optical power of the near-IR excitation beam (first order of deflection from the AOM) is controlled by adjusting the RF signal amplitude, and it is focused onto the membrane by using a 75-mm lens (L2). \\ On the other side (right side in Fig.~\ref{fig:exp_setup}), the membrane is illuminated by the MIR QCL probe beam. The QCL is operated at room temperature (\SI{18}{ \celsius }) in a single-mode regime at a bias current of \SI{470}{mA}. The laser mounting arrangement allows us to conveniently exploit the emission from both laser facets. The front-facet-emitted beam is used to probe the membrane oscillation, receiving back the resulting optical feedback. To this extent, this radiation is focused on the membrane via a 50-mm lens (L1), and a germanium wedged window (W) is used in the probe path to prevent any injection of the near-IR radiation into the QCL. Instead, the radiation emitted by the back facet of the QCL can be used for free-space transmission of the MIR radiation carrying the SM signal. In our setup, after a free-space propagation, the SM signal is retrieved by detecting the back-facet-emitted radiation via a fast commercial photovoltaic detector (PVI-4TE-5-2x2 by Vigo System). The detector output signal is used as input signal of a lock-in amplifier. Depending on the application, we highlight that the direct detection of the SM signal from the laser voltage variation could be more convenient, as this configuration requires no extra detector. In particular, the presented optomechanical platform can be conveniently exploited in both configurations~\cite{Vezio:2025}.
\\As depicted in Fig.~\ref{fig:exp_setup}, our system can be operated in two different configurations. In the first configuration (\textit{Conf. 1}), the PZT induces the membrane oscillation. This configuration is used for system calibration, as it allows us to quantify the photothermal effect on the membrane of both pump and probe under quasi-continuous-wave operation (see Sec.~\ref{sec:Results}). In the second configuration (\textit{Conf. 2}), the membrane oscillation is induced by AM modulating the excitation beam (without PZT actuation), as discussed in Section~\ref{sec:Results}. In both cases, the lock-in amplifier is used: 1) to control the sinusoidal modulation signal and its frequency sweep, which is sent to the PZT in \textit{Conf. 1} or to the AOM in \textit{Conf. 2}; 2) to demodulate the self-mixing signal measured via the photodetector. This allows the retrieval of the membrane spectral response (e.g., as shown in Fig.~\ref{fig:freq_shift_pzt_QCL}a). \\In both configurations, the impinging power onto the photovoltaic detector has been kept constant, and the detector is operated in the linear responsivity regime as in refs. \cite{Gabbrielli:2021, Gabbrielli:2025}. 
\section{Results and Discussion}
\label{sec:Results}
\begin{figure*}[htb!]
    \centering
\includegraphics[width=0.95\linewidth]{ 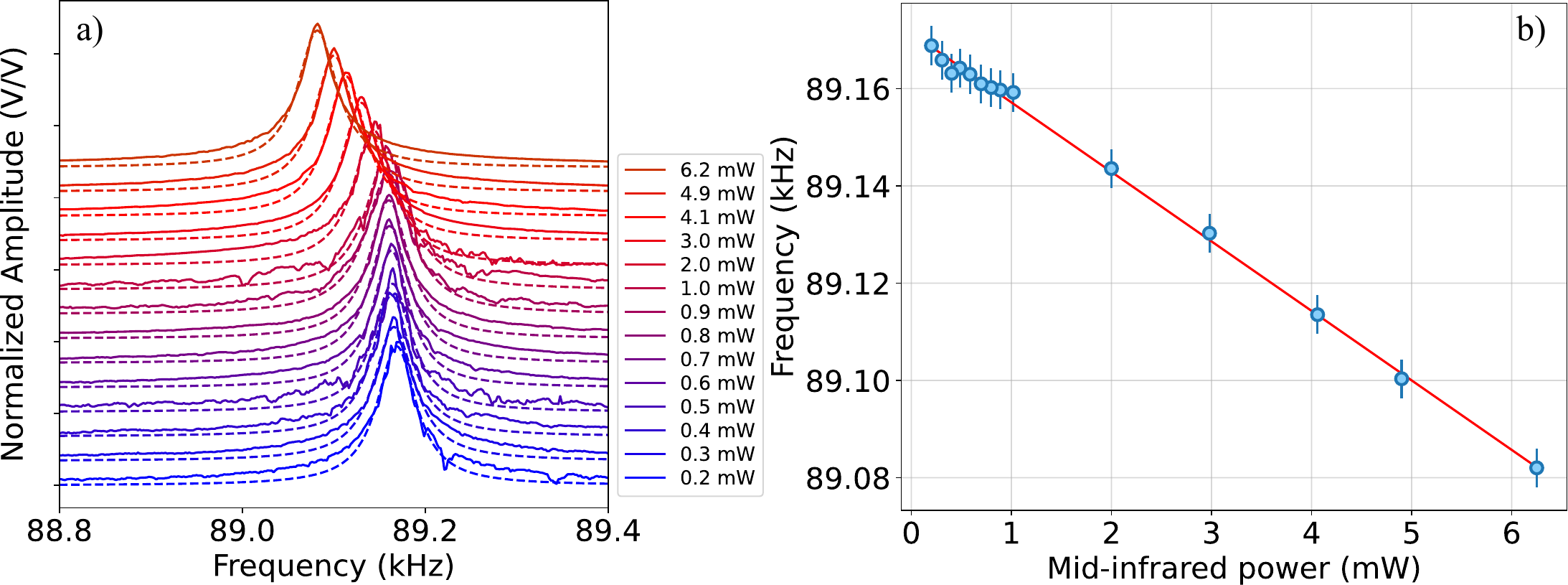}
    \caption{a): Amplitude peak normalized to its maximum value as a function of PZT modulation frequency, plotted at different values of probing mid-IR optical power, when the membrane oscillation is excited by the PZT and no excitation light is sent onto the membrane. Each peak is fit with a Lorentzian function, allowing the estimation of the resonance frequency at each impinging power value reported in the legend. b) Membrane resonance frequency, estimated via the Lorentzian fit, as a function of mid-IR probing power. The data (blue points) are fit with a straight line (red curve). The error bar (blue lines) is calculated as the standard deviation of the repeated measurements of the resonance frequency for a certain power value.}
    \label{fig:freq_shift_pzt_QCL}
\end{figure*}
At first, the frequency response of the optomechanical system is studied when only the probing radiation, i.e., the QCL's one, is present and using the PZT to drive the membrane oscillation (\textit{Conf}. 1 in Fig.~\ref{fig:exp_setup}). In detail, we study the resonance frequency shift while varying the QCL power (Fig.~\ref{fig:freq_shift_pzt_QCL}). To this extent, the QCL is kept at a constant working condition (bias current \SI{470}{mA} and temperature \SI{18}{\celsius}), and a variable attenuator is used to control the power impinging on the membrane. 
As shown in Fig.~\ref{fig:freq_shift_pzt_QCL}a, a redshift of the resonance frequency (from \SI{89.082}{kHz} to \SI{89.169}{kHz}) occurs due to thermal effects when increasing the mid-IR optical power (from \SI{0.2}{mW} to \SI{6.2}{mW}). Indeed, the metal surface of the membrane absorbs the infrared light, resulting in a relaxation of the mechanical stress on the trampoline membrane. For each power value, the resonance frequency is estimated by fitting each resonance peak of Fig.~\ref{fig:freq_shift_pzt_QCL}a with a Lorentzian function. Fig.~\ref{fig:freq_shift_pzt_QCL}b illustrates the linear trend of the frequency shift against the impinging QCL power. While increasing the impinging power, the peak frequency decreases at a rate of $ (14.28 \pm 0.17) $~Hz/mW as estimated via linear fit (red line).
\begin{figure*} [ht!b]
    \centering
    \includegraphics[width=0.95\linewidth]{ 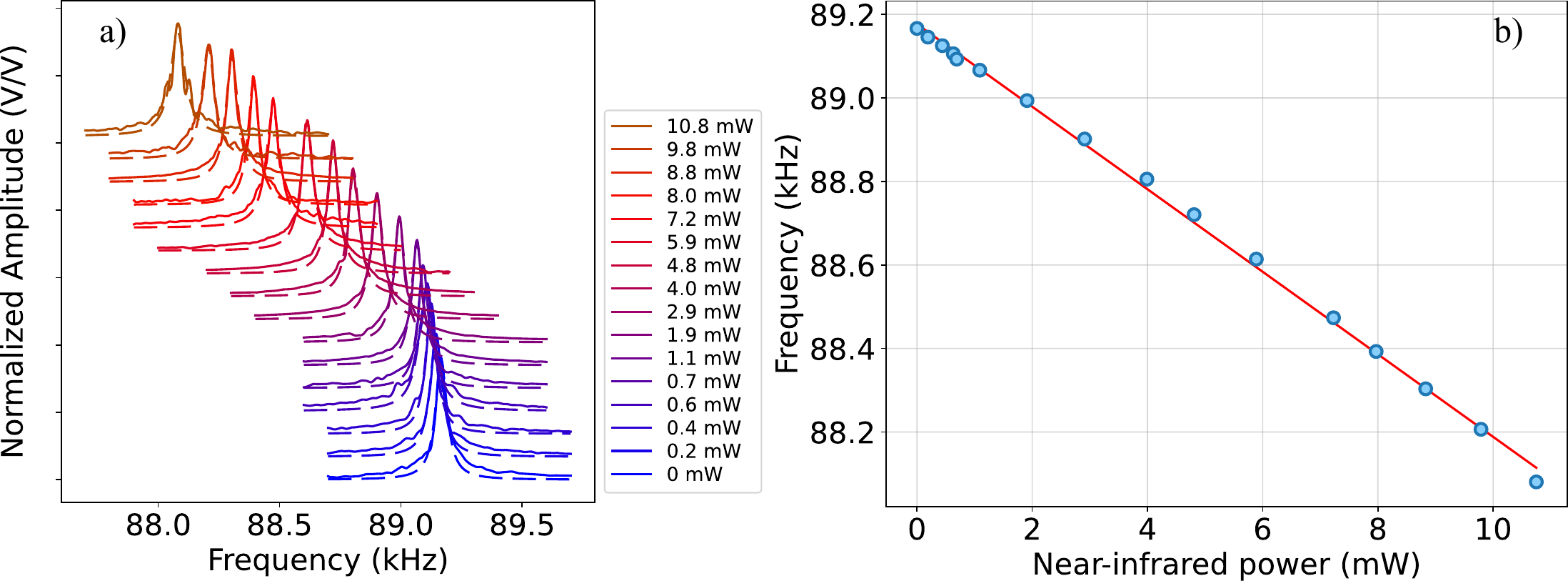}
    \caption{a): Normalized amplitude peak to its maximum value as a function of PZT modulation frequency, plotted at different values of near-IR impinging power, when the membrane oscillation is driven via the PZT and the mid-IR probe beam is kept at a fixed working condition, emitting a power of \SI{6.2}{mW}. Each peak is fitted via a Lorentzian function, which allows us to estimate the resonance frequency at the peak value for the different values of impinging power. The obtained values are depicted as blue points in graph (b). b): Membrane resonance frequency as a function of near-IR impinging power. The data (blue points) are fitted via a linear fit (red line). Here, the error bars, obtained as in Fig.~\ref{fig:freq_shift_pzt_QCL}, are not visible.}
    \label{fig:resonance_with_QCL_nearIR_and_PZT}
\end{figure*}
After this preliminary characterization, we test the membrane response when illuminated with both the mid-IR probe beam and near-IR excitation beam (Fig.~\ref{fig:resonance_with_QCL_nearIR_and_PZT}). For this test, the membrane oscillation is driven by the PZT controlled via the lock-in amplifier (\textit{Conf}.~1 in Fig.~\ref{fig:exp_setup}).
The mid-IR light impinging on the membrane is fixed at its maximum, i.e. \SI{6.2}{mW} while the excitation beam power is tuned by changing the amplitude of the AOM RF modulation via the signal generator (see Fig.~\ref{fig:exp_setup}). At this probe power level, the feedback signal is strong, and, at the same time, the probe radiation does not saturate the thermal effects on the membrane, as demonstrated by the redshift of the resonance frequency induced by the near-IR radiation (Fig.~\ref{fig:resonance_with_QCL_nearIR_and_PZT}a). Therefore, an independent analysis of the near-IR laser effect on the membrane resonance is possible.
As clearly visible in Fig.~\ref{fig:resonance_with_QCL_nearIR_and_PZT}a, the resonance peak is redshifted in frequency when the near-IR laser power increases. Again, for each value of the near-IR impinging power, we estimate the membrane resonance frequency via a Lorentzian fit of the experimental data. The results are shown in Fig.~\ref{fig:resonance_with_QCL_nearIR_and_PZT}b. The resonance frequency (blue points) linearly decreases as the near-IR power increases at a rate of $(98.8 \pm 1.1)$ Hz/mW, as estimated via a linear fit (red line). We remark that the difference between this rate value and the one related to the QCL power variation (Fig.~\ref{fig:freq_shift_pzt_QCL}) can be explained by different factors: mode matching; the direction of the incident beams on the membrane; the use of different windows in the two paths (transmissivity of CaF$_2$ windows at \SI{4.5}{\micro m}: 95\%, transmissivity of COC window at \SI{1.064}{\micro m}: 85\%); and the fact that, on the 1064 nm side, absorption is also influenced by the presence of a 3-nm Cr layer between the SiN and the 50-nm Au layer~\cite{Vezio:2025}.
\begin{figure*} [htb!]
    \centering
\includegraphics[width=0.95\linewidth]{ 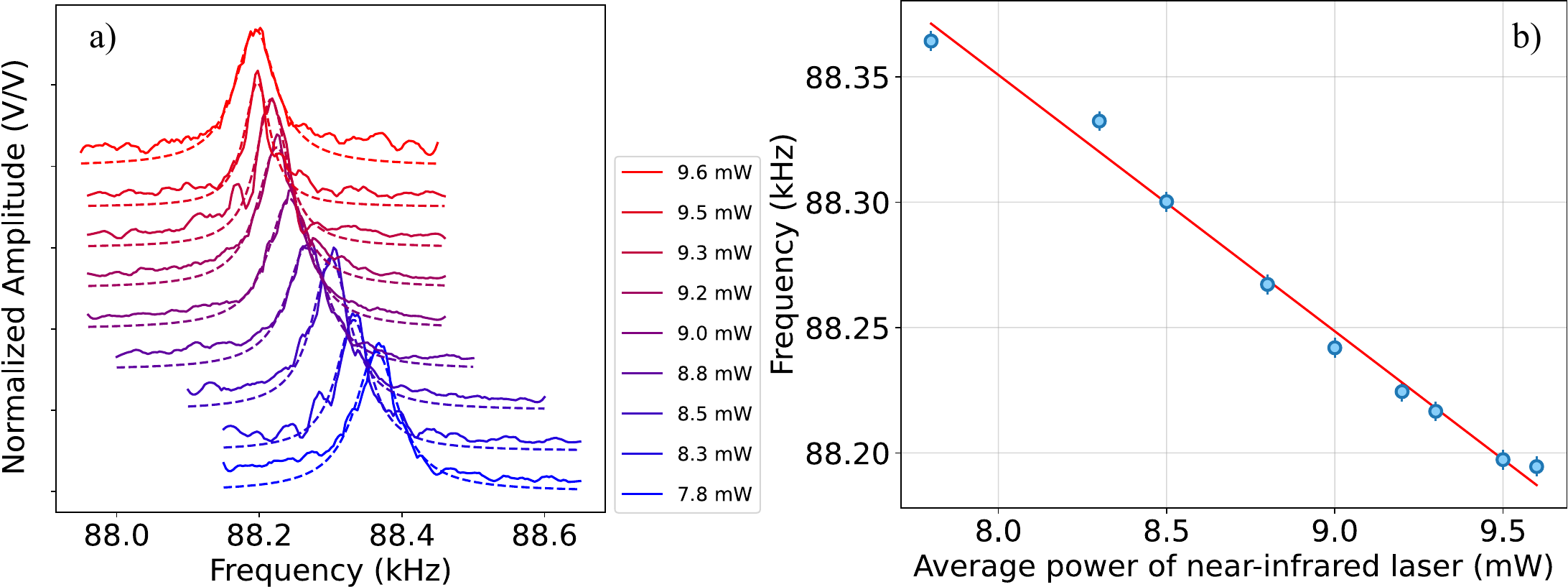}
    \caption{a): Amplitude peak normalized to its maximum value as a function of AM modulation frequency, plotted at different values of AM-modulated near-IR impinging power, when the mid-IR probe beam is kept at a fixed working condition at a power of \SI{6.2}{mW}. Each peak is fitted via a Lorentzian function, which allows us to estimate the resonance frequency at the peak value for the different values of impinging power. The obtained values are depicted as blue points in graph (b). b): Membrane resonance frequency as a function of
near-IR impinging power. The data (blue points) are fit to a straight line (red curve). The error bars are obtained as in Fig.~\ref{fig:freq_shift_pzt_QCL}. } \label{fig:resonance_peak_AM_mod_NIR}
\end{figure*}

Finally, we test the membrane response in \textit{Conf.~2}. Specifically, in this configuration, the AM-modulated near-IR beam induces the membrane oscillation, which is probed via the mid-IR beam kept at constant working conditions with a fixed optical power of~\SI{6.2}{mW} (Fig.~\ref{fig:resonance_peak_AM_mod_NIR}). As described in section~\ref{sec:Exp_setup}, the AM modulation of the near-IR light is obtained by externally driving the AOM via the lock-in amplifier. Moreover, to reconstruct the resonance peak, the AM modulation is swept in the frequency range of the membrane resonance. The procedure is repeated for different values of AM modulation amplitude. This allows us to monitor the frequency shift of the membrane resonance peak for different values of the near-IR excitation beam power. As shown in Figs.~\ref{fig:resonance_peak_AM_mod_NIR}a and \ref{fig:resonance_peak_AM_mod_NIR}b, a blue-shift is visible in the peak frequency when increasing the AM modulation amplitude from \SI{100}{mV} to \SI{500}{mV}, corresponding to an average near-IR power from \SI{9.6}{mW} to \SI{7.8}{mW},
respectively. In particular, we can see that, in this case, the frequency shift follows a linear trend when plotted against the impinging near-IR power with a shift rate of $(102 \pm 4)$~Hz/mW. 
In this second configuration, the membrane oscillation, i.e., the presence of the resonance peak in Fig.~\ref{fig:resonance_peak_AM_mod_NIR}, is due to light-induced force exerted by the near-IR beam. Moreover, the increase in impinging power leads to a redshift of the resonance frequency, as already shown also in \textit{Conf. 1}. In general, the light-induced force might have different contributions, i.e., radiation pressure and photothermal force \cite{Metzger:2008}. Typically, these two phenomena have different time scales, i.e., photothermal force is characterized by a slower time scale compared to the quasi-instantaneousness of the radiation pressure \cite{Metzger:2008}. However, even if their time scale can be quite different depending on the specific oscillating system (e.g., membrane and substrate materials), the thermal effect can be non-negligible—even if dumped— at the resonance frequency\cite{Vezio:2025,Metzger:2008}. 
\begin{figure} [!htb]
    \centering
\includegraphics[width=0.8\linewidth]{ 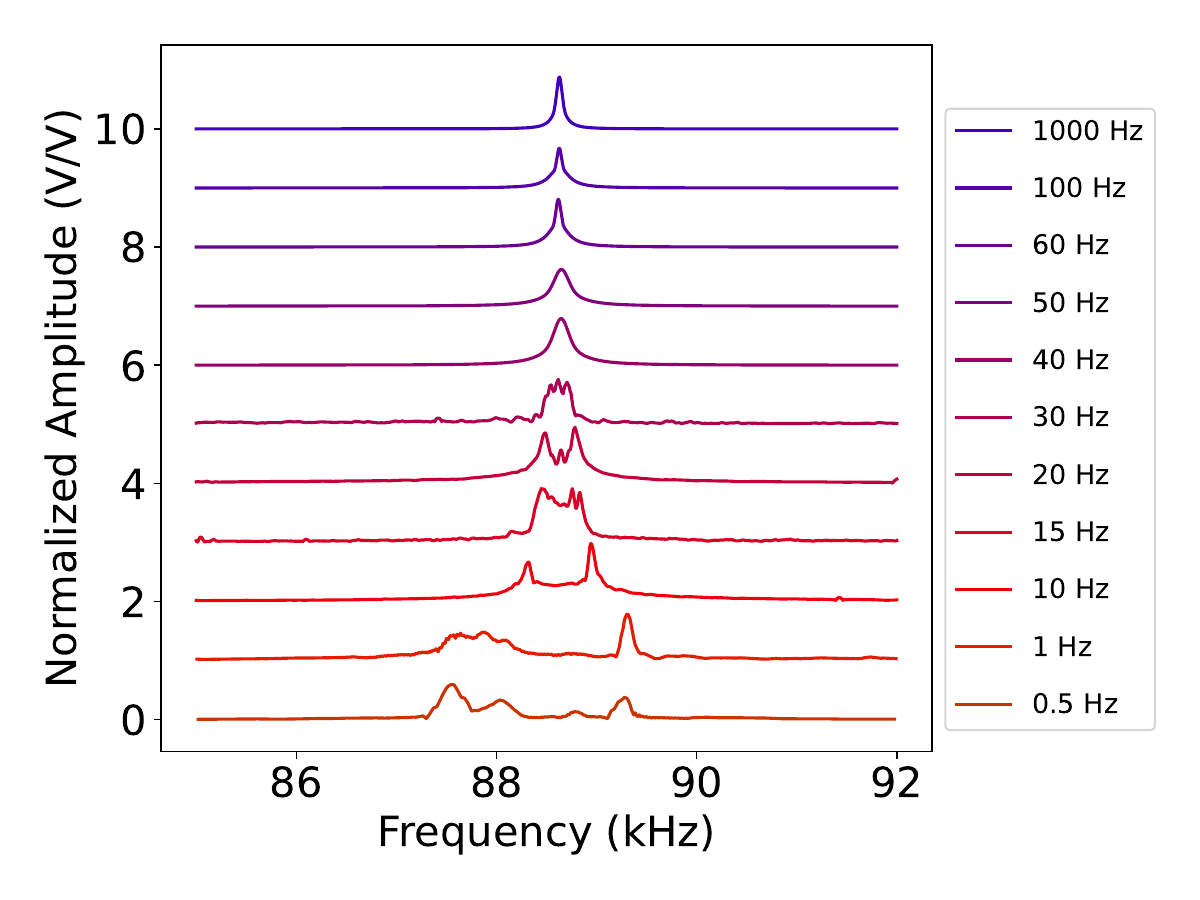}
    \caption{Characterization of the thermal relaxation time scale onto the membrane peak of resonance when the system is driven in \textit{Conf.1} and a slow AM modulation (see the legend) is added on the excitation beam. }
    \label{fig:thermal_eff}
\end{figure}
Therefore, depending on the applications and specific setup, the driving optomechanical force of the membrane oscillation can be dominated by radiation pressure, photothermal force, or a combination of the two. In view of further investigation to understand the {photo-thermal contribution} in the system, we report a preliminary characterization of its time scale. {The SM setup is set in \textit{Conf. 1}, i.e. the membrane is oscillated using the PZT driven with the lock-in amplifier while impinging onto the membrane with both the excitation and probe beams}. Moreover, we externally apply a slow AM modulation (as reported in the legend of Fig.~\ref{fig:thermal_eff}) to the excitation beam by driving the AOM with the signal generator. If the AM modulation is slow enough (i.e., comparable to the thermal relaxation time scale), two peaks are appreciable in the spectrum, as shown in Fig.~\ref{fig:thermal_eff}. The first one is at the resonance frequency of the membrane when illuminated with just the probe radiation (no excitation beam). In contrast, the second peak, which is redshifted compared to the previous one, is due to the thermal effect of illuminating the membrane with both the excitation and probe radiation. The net effect of AM modulating the excitation beam is a frequency modulation (FM) of the resonance peak. The spectra are, in fact, characterized by the typical shape of a frequency-modulated signal~\cite{Corrias:2022}. The spectrum is made of two clearly separated peaks up to \SI{20}{Hz}, while starting from \SI{40}{Hz}, and above, they merge into one single peak. This implies that the thermal relaxation effect bandwidth is lower than \SI{40}{Hz} {and significantly below the membrane resonance frequency ($\approx \SI{90}{kHz}$). {This suggests a significant dump of thermal effect in the time scale characterizing the membrane oscillation; however, deeper investigations are needed to understand which is the main driving force.}}
To summarize, by AM modulating the excitation radiation in the range of the membrane resonance frequency, we demonstrate the capability of encoding the AM modulation information into the SM  signal in the QCL light when the membrane is excited via light-induced force. Therefore, the presented optomechanical setup promises to serve as an information transducer between the two different wavelengths.
\section{Conclusion and Perspectives}
\label{sec:ConclusionandPresp}
This work describes an optomechanical system capable of transferring information from the near-IR to the mid-IR. In detail, an SM-assisted transduction scheme is presented, where a mid-IR QCL is used in a laser-feedback interferometer to probe the oscillation of a membrane when excited via a near-IR beam.
{ This enables the monitoring of membrane oscillations $-$ fully optically induced by the near-IR beam $-$ through the SM signal when the beam is amplitude-modulated at frequencies close to the membrane resonance.} Therefore, by demonstrating the capability of detecting the resonance signal, we show the potential of using this optomechanical system as a communication gate (i.e., information transducer) between two different spectral regions. Indeed, the oscillation of the membrane can be used to amplify and transmit the AM modulation signal of the excitation beam to the probe.  
The tested membrane interface is wavelength-independent, therefore allowing a connection between the mid-IR and the excitation beam, regardless of its color.  Moreover, in principle, this system can also be exploited to control several QCL parameters, like frequency, intensity, phase, or power, by SM, which boasts a bandwidth close to 100 kHz. \\  {From a technical perspective, this system can be exploited to control the QCL laser operation via its SM coupling with an external optomechanical system, benefiting from its faster mechanical dynamics as compared to bulk components. \\ From a more generic perspective, by engineering optimized membranes for communication purposes (e.g., high resonance frequencies), the techniques presented here could be exploited to create optomechanical interfaces between the fiber-based telecom communication infrastructure and mid-IR free-space communication channels. The latter are already used for point-to-point communication, also overcoming adverse weather conditions due to the much reduced scattering\cite{Corrias:2022,Seminara:22,Su:18, Flannigan_2022}. Additionally, starting from this optomechanical platform, novel sensing and imaging setups can be developed: a membrane-based array, for instance, could allow spatial control of the modulation amplitude in the probe, either exploiting the pump intensity distribution or the different mechanical frequencies of each membrane \cite{gregorat2024highly}. Furthermore, this system can pave the way to advanced imaging methods where the mid-IR laser SM signal could be used to probe, e.g, near-IR or visible photothermally excited substrates or targets, envisioning hybrid photoacoustic mid-IR (or THz) sensing \cite{Pelini2025}.}

\paragraph{Data Availability Statement}

The data that support the findings of this study is available from the
corresponding author upon reasonable
request.

\begin{acknowledgement}

The author would like to thank Irene La Penna (LENS) for her help in pre-aligning the \SI{1}{\micro m } laser from the laser output to the AOM, Davide Mazzotti (CNR-INO) for the fruitful conversation about possible perspectives, and Mario Siciliani De Cumis (ASI) {and Francesco Marino (CNR-INO)} for the useful discussion about thermal effects.
\newline
The authors acknowledge financial support from the
European Union—Next Generation EU with the Italian National Recovery and Resilience Plan (NRRP), Mission 4, Component 2, Investment 1.3, CUP D43C22003080001, partnership on “Telecommunications of the Future” (PE00000001 — program “RESTART”), and the I-PHOQS Infrastructure ”Integrated infrastructure initiative in Photonic and Quantum Sciences” [IR0000016, ID D2B8D520], with the Laserlab-Europe Project [G.A. n.871124], with the MUQUABIS Project “Multiscale quantum bio-imaging and spectroscopy” [G.A. n.101070546], with the QUID project "Quantum Italy Deployment" [G.A. No 101091408]; from the Italian ESFRI Roadmap (Extreme Light Infrastructure - ELI Project); from ASI and CNR under the Joint Project “Laboratori congiunti ASI-CNR nel settore delle Quantum Technologies (QASINO)” (Accordo Attuativo n. 2023-47-HH.0); from the Italian Ministero dell'Università e della Ricerca (project PRIN-2022KH2KMT QUAQK). 

\end{acknowledgement}

\bibliography{ref}

\end{document}